\documentclass[sigconf]{acmart}

\AtBeginDocument{%
  \providecommand\BibTeX{{%
    \normalfont B\kern-0.5em{\scshape i\kern-0.25em b}\kern-0.8em\TeX}}}

\setcopyright{acmcopyright}
\copyrightyear{2022}
\acmYear{2022}
\acmDOI{XXXXXXX.XXXXXXX}

\usepackage{caption}
\usepackage{subcaption}

\acmConference[EICC '22]{European Interdisciplinary Cybersecurity Conference}{June 15--16,
  2022}{Barcelona, Spain}
\acmBooktitle{European Interdisciplinary Cybersecurity Conference,
 June 15--16, 2022, Barcelona, Spain} 
\acmPrice{15.00}
\acmISBN{978-1-4503-XXXX-X/18/06}

\usepackage{algorithm}
\usepackage[]{algpseudocode}
\usepackage{multirow}

\algnewcommand\algorithmicforeach{\textbf{for each}}
\algdef{S}[FOR]{ForEach}[1]{\algorithmicforeach\ #1\ \algorithmicdo}
\newcommand{\LineComment}[1]{// \textit{#1} }

\begin{document}

\title{Homomorphic Encryption and Federated Learning based Privacy-Preserving CNN Training: COVID-19 Detection Use-Case}


\author{Febrianti Wibawa}\authornotemark[1]
\email{f.febrianti@gmail.com}
\author{Ferhat Ozgur Catak}
\authornotemark[2]
\email{f.ozgur.catak@uis.no}
\affiliation{%
  \institution{Department of Electrical Engineering and Computer Science, University of Stavanger}
  \city{Stavanger}
  \state{Rogaland}
  \country{Norway}
  \postcode{4021}
}

\author{Salih Sarp}
\email{sarps@vcu.edu}
\affiliation{%
  \institution{Department of Electrical and Computer Engineering, Virginia Commonwealth University}
  \streetaddress{601 West Main Street, Room 203
Box 843072}
  \city{Richmond}
  \state{Virginia}
  \country{USA}
  \postcode{23284}
}

\author{Murat Kuzlu}
\email{mkuzlu@odu.edu}
\affiliation{%
  \institution{Batten College of Engineering and Technology, Old Dominion University}
  \streetaddress{211B KAUFMAN HALL}
  \city{Norfolk}
  \state{Virginia}
  \country{USA}
  \postcode{23529}
}

\author{Umit Cali}
\email{umit.cali@ntnu.no}
\affiliation{%
  \institution{Norwegian University of Science and Technology}
  \streetaddress{}
  \city{Trondheim}
  \state{}
  \country{Norway}
  \postcode{}
}

\renewcommand{\shortauthors}{Wibawa and Catak, et al.}


\begin{abstract}
  Medical data is often highly sensitive in terms of data privacy and security concerns. Federated learning, one type of machine learning techniques, has been started to use for the improvement of the privacy and security of medical data. In the federated learning, the training data is distributed across multiple machines, and the learning process is performed in a collaborative manner. There are several privacy attacks on deep learning (DL) models to get the sensitive information by attackers. Therefore, the DL model itself should be protected from the adversarial attack, especially for applications using medical data. One of the solutions for this problem is homomorphic encryption-based model protection from the adversary collaborator. This paper proposes a privacy-preserving federated learning algorithm for medical data using homomorphic encryption. The proposed algorithm uses a secure multi-party computation protocol to protect the deep learning model from the adversaries. In this study, the proposed algorithm using a real-world medical dataset is evaluated in terms of the model performance.
\end{abstract}




\maketitle

\section{Introduction}
Machine learning (ML) is a widely used technique in almost all fields, where a computer system can learn from data to improve its performance. This technique is widely used in many application areas such as image recognition, natural language processing, and machine translation. Federated learning is a machine learning technique where the training data is distributed across multiple machines, and the learning process is performed in a collaborative manner \cite{kairouz2021advances}. This technique can be used to improve the privacy and security of medical data \cite{csahinbacs2021secure}.

Medical data is often highly sensitive and is often subject to data privacy and security concerns \cite{abouelmehdi2017big}. For example, a person's health information is often confidential and can be used to identify the person. Thus it is essential to protect the privacy and security of medical data. Health Insurance Portability and Accountability Act (HIPAA) (US Department of Health and Human Services, 2014) and General Data Protection Regulation (GDPR) (The European Union ,2018) strictly mandate the personal health information privacy. There are various methods to safeguard the private information.  Federated learning is one of the techniques that can be utilized for the protection of sensitive data during multi-party computation tasks. This technique can be used to improve the privacy and security of medical data by preventing the data from being centralized and vulnerable.

Keeping the data local is not sufficient for the security of the data and the ML model. However, there are several privacy attacks on deep learning models to get the private data \cite{electronics9020229,catak2018}.  For example, the attackers can use the gradient information of the deep learning model to get the sensitive information. Thus the deep learning model itself should be protected from the adversaries as well. One of the solutions for this problem is homomorphic encryption-based model protection from the adversary collaborator. Homomorphic encryption is a technique where the data can be encrypted, and the operations can be performed on the encrypted data \cite{alloghani2019systematic}. This technique can be used to protect the deep learning model from the adversaries.

This paper proposes a privacy-preserving federated learning algorithm based convolutional neural network (CNN) for medical data using homomorphic encryption. The proposed algorithm uses a secure multi-party computation protocol to protect the deep learning model from the adversaries. We evaluate the proposed algorithm using a real-world medical dataset and show that the proposed algorithm can protect the deep learning model from the adversaries.


\section{Related Work}

Data-driven ML models provide unprecedented opportunities for healthcare with the use of sensitive health data. These models are trained locally to protect the sensitive health data. However, it is difficult to build robust models without diverse and large datasets utilizing the full spectrum of health concerns. Prior proposed works to overcome this problems include federated learning techniques. For instance, the studies \cite{xu2021federated, rieke2020future, antunes2022federated} reviewed the current applications and technical considerations of the federated learning technique to preserve the sensitive biomedical data. Impact of the federated learning is examined through the stakeholders such as patients, clinicians, healthcare facilities and manufacturers. In another study, the authors in \cite{li2019privacy} utilized federated learning systems for brain tumour segmentation on the BraTS dataset which consist of magnetic resonance imaging brain scans. The results show that performance is decreased by the privacy protection costs. Same BraTS dataset is used in \cite{sheller2018multi} to compare three collaborative training techniques, i.e., federated learning, institutional incremental learning (IIL) and cyclic institutional learning (CIIL). In IIL and CIIL, institutions train a shared model successively where CIIL adds a cycling loop through organisations. The results indicates that federated learning achieves similar Dice scores to that of models trained by sharing data. It outperform the IIL and CIIL methods since these methods suffer from catastrophic forgetting and complexity. 

Medical data is also safeguarded by encryption techniques such as homomorphic encryption. In \cite{kumar2020secure}, authors propose an online secure multiparty computation with sharing patient information to hospitals using homomorphic encryption. Bocu et al. \cite{bocu2018homomorphic} proposed a homomorphic encryption model that is integrated to personal health information system utilizing heart rate data. The results indicates that the described technique successfully addressed the requirements for the secure data processing for the 500 patients with expected storage and network challenges. In another study by Wang et al. \cite{wang2015data} proposed a data division scheme based homomorphic encryption for wireless sensor networks. The results show that there is trade off between resources and data security. In \cite{kara2021fully}, applicability of homomorphic encryption is shown by measuring the vitals of the patients with a lightweight encryption scheme. Sensor data such as respiration and heart rate are encrypted using homomorphic encryption before transmitting to the non-trusting third party while encryption takes place only in medical facility. The study in \cite{talpur2015shared} developed an IoT based architecture with homomorphic encryption to combat data loss and spoofing attacks for chronic disease monitoring. results suggest that homomorphic encryption provide cost effective and straightforward protection to the sensitive health information. Blockchain technologies are also utilized in cooperation with homomorphic encryption for the security of medical data. Authors in \cite{tan2020practical} proposed a practical pandemic infection tracking using homomorphic encryption and blockchain technologies in intelligent trasnportatiton systems using automatic healthcare monitoring. In another study Ali et al. \cite{ali2022deep} developed a search-able distributed medical database on a blockchain using homomorphic encryption. 
The increase need to secure sensitive information leads to use of various techniques together. In the scope of this study, a multi-party computation tool using federated learning with homomorphic encryption is developed and analyzed.


\section{Preliminaries}
\subsection{Homomorphic encryption}

Nowadays data encryption is a common practice not only for enterprises but also individuals. It is meant to protect privacy of the data. Data encryption mostly done at rest, when the data is stored and in transit when the data is transferred. However data encryption is not popularly used upon when running or executing the operations or computations. 

Homomorphic encryption is an encryption method which allows arithmetical computations to be performed directly on encrypted or ciphered text without requiring any decryption. Outputs of the computations are also in encrypted form and provide identical or almost identical result when decrypted. This means that Homomorphic encryption allows data processing without disclosing the actual data.

If $Enc$ denotes encryption, $Dec$ denotes decryption,
and $f()$ is a function applied on actual values (plaintexts) $a$ and $b$, using encrpytion key $pk$ , then homomorphic encryption property would be: \\

\centerline{$f(a, b) = Dec(Enc(pk,a), Enc(pk,b))$}

Homomorphic encryption can be used for privacy-preserving outsourced storage and computation. This allows data to be encrypted and out-sourced to commercial cloud environments for processing, all while encrypted. 


\par There are several types of homomorphic encryption \cite{acar2018survey};
\begin{enumerate}
  \item Partially homomorphic encryption is homomorphic encryption that supports only one homomorphic operation, either  addition or multiplication, with unlimited number of times.
  
  \item Somewhat homomorphic encryption schemes allows both addition and multiplication but only in a limited number of times.
  
  \item Leveled fully homomorphic encryption supports the evaluation of arbitrary circuits composed of multiple types of gates of bounded (pre-determined) depth.
  
  \item Fully homomorphic encryption (FHE) supports both addition and multiplication operations with unlimited number of times.
\end{enumerate}

Somewhat homomorphic encryption (SHE) is used in this work since it allows both addition and multiplication operations on encrypted data which is required in aggregation of machine learning model weights.
 
\subsection{Brakerski-Fan-Vercauteren (BFV) scheme}
The BFV scheme is a well-known homomorphic encryption scheme. It encrypts polynomials instead of bits. The encrypted polynomials can be evaluated homomorphically. It is secure in the sense that it is CCA secure. The security is based on the hardness of the problem SIS. It can be described as follows.

We now briefly describe the BFV scheme. Let $n$ be a positive integer, $q$ be a prime number, $\mathbb{F}_{q}$ be the finite field with $q$ elements, $t$ be a positive integer, $(\alpha_{0},\alpha_{1},\ldots,\alpha_{n-1})$ be a random tuple in $\mathbb{F}_{q}^{n}$, $s$ be a positive integer, $\eta$ be a positive integer. Let $N=q^s$ and $M=q^{\eta}$. The secret key is $(\alpha_{0},\alpha_{1},\ldots,\alpha_{n-1})$. The public key is $(\alpha_{0}^{N},\alpha_{1}^{N},\ldots,\alpha_{n-1}^{N})$. The message space is $\mathbb{F}_{q}[x]_{<t}$. The message $m(x)$ is encrypted to $c(x)=\frac{m(x)}{(x-\alpha_{0})(x-\alpha_{1})\ldots(x-\alpha_{n-1})}+e(x)$, where $e(x)$ is a polynomial of degree less than $t$. The decryption is done by evaluating $c(x)$ at all points of the form $\alpha_{i}^{M}$ and then interpolating $m(x)$ from the resulting evaluations.

\subsection{Regulatory Aspects of Privacy in Health Sector}
Trust and privacy are among the fundamental elements of digital healthcare systems and platforms. The trust is expected to be built between various stakeholders of the digital healthcare ecosystems such as patients, medical care providers, health authorities and healthcare systems providers. The following medical data are among the most critical ones in terms of privacy and have to be protected: 

\begin{itemize}
    \item Personal information related to patient such as address, social security number, birth date, and bank account number,
    \item Provided medical and psychological services, drugs, equipment, and procedures,
    \item Status of the patients’ medical or psychological conditions,
    \item The information related to the hospital, clinic or the medical professionals who provided the medical and psychological services.
   
\end{itemize}
The European General Data Protection Regulation (GDPR) is among the mostly applied regulatory framework in terms of data privacy that concentrates on individual control for data subjects of ‘their’ data. Public and private healthcare data privacy is handled under GDPR regulations \cite{VANVEEN201870}.

\subsection{BFV Scheme}
The Brakerski/Fan-Vercauteren (BFV) architecture \cite{10.1007/978-3-642-32009-5_50,Fan2012SomewhatPF,EURECOM_6801} incorporates powerful Single Instruction Multiple Data (SIMD) parallelism, making it ideal for applications that handle massive volumes of data. In this crypto scheme, the messages are the vectors of integers, $\mathbf{m} \in \mathbb{Z}^n$. The messages are encoded into plaintext polynomials of degree $n$. 

\subsection{Federated Learning}
Federated learning is a machine learning technique that enables multiple parties to build and train a common machine learning model without exchanging or sharing data. Each party (client) stores and processes their own dataset (local dataset) while there is a common model shared with all parties (clients). In this case each client trains the common model using local dataset, and sends trained model to a centralized server. The server then aggregates model received from all the clients and distributes the aggregated model back to the clients. 

Federated learning addresses data security and privacy issues since it doesn't require access to dataset of each client, nor requires the dataset to be distributed. The local dataset itself doesn't have to be identically distributed and can be heterogeneous. This behaviour makes Federated Learning more popular in healthcare applications. Federated Learning enables health institutions to form and train a common model without transferring sensitive patient data out.

There are several types of Federated Learning setting:\cite{barbieri2022decentralized}
\begin{enumerate}
    \item Centralized federated learning. In this setting, a central server is used to populate and aggregate models from participating clients during learning process. A global common model is pushed from the server down to the clients.

    \item Decentralized federated learning. In this setting, participating clients coordinate among themselves to obtain a global common model \cite{roy2019braintorrent}. 
    
    \item Heterogeneous federated learning. In this setting, participating clients come from different technical platfrom, e.g. PC and mobile phones, with own local dataset and model while obtaining single global model.
    
\end{enumerate}

In this work, centralized federated learning setting is implemented, to demonstrated model aggregation by single centralized server.

\section{System Model}

\begin{figure*}[!htbp]
    \centering
    \includegraphics[width=0.8\linewidth]{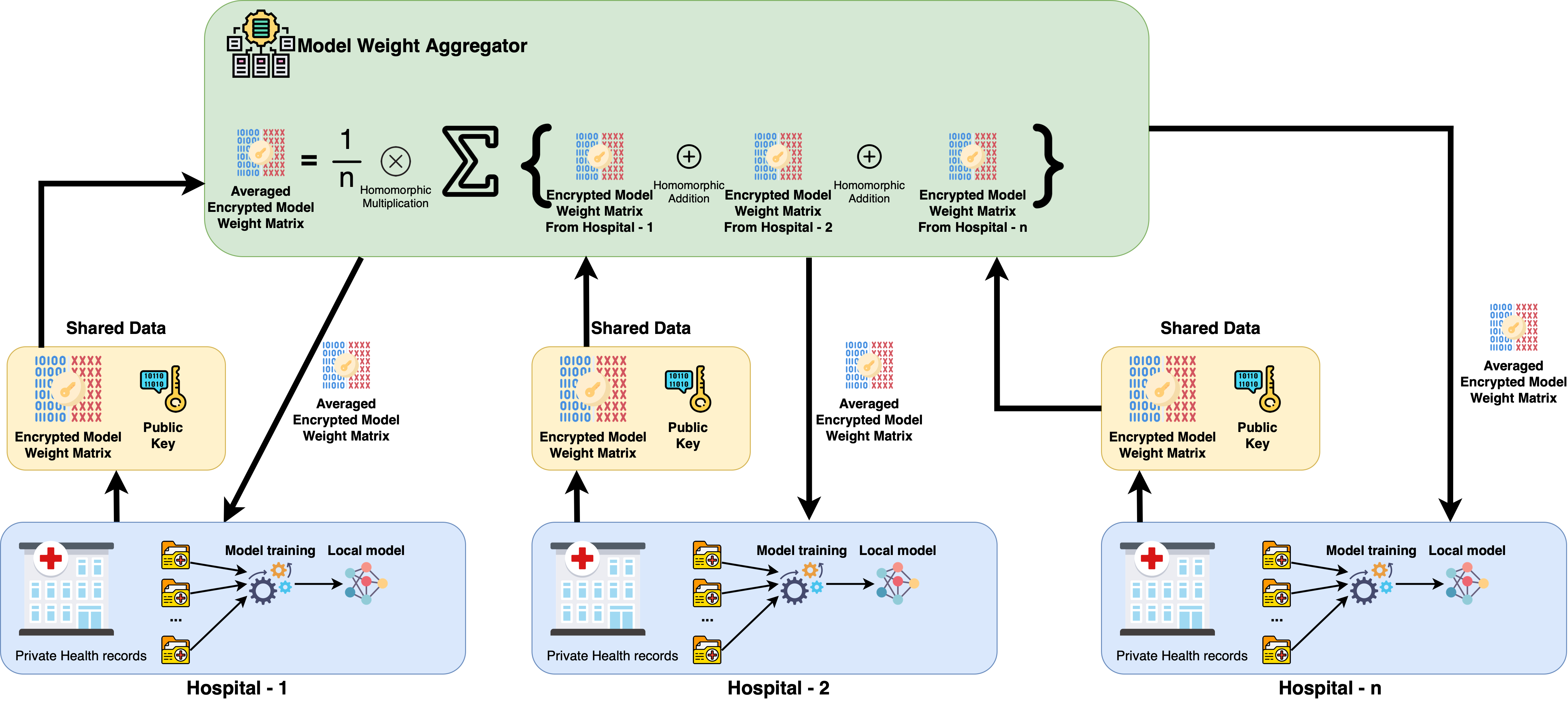}
    \caption{Overall system overview of the proposed method}
    \label{fig:system_overview}
\end{figure*}

This section gives a high-level system overview of the proposed BFV crypto-scheme-based privacy-preserving federated learning COVID-19 detection training method. The proposed privacy-preserving scheme is a two-phase approach: (1) local model training at each client and (2) encrypted model weight aggregation at the server. In the local model training phase, each client builds their local CNN based DL model using their local electronic health record dataset. The clients encrypt the model weights matrix using the public key. In the second step, the server aggregates all clients' encrypted weight matrices and sends the final matrix to the clients. Each client decrypts the aggregated encrypted weight matrix to update the model weights of their DL model. Figure \ref{fig:system_overview} shows the system overview.

\begin{figure}[htbp!]
    \centering
    \includegraphics[width=0.9\linewidth]{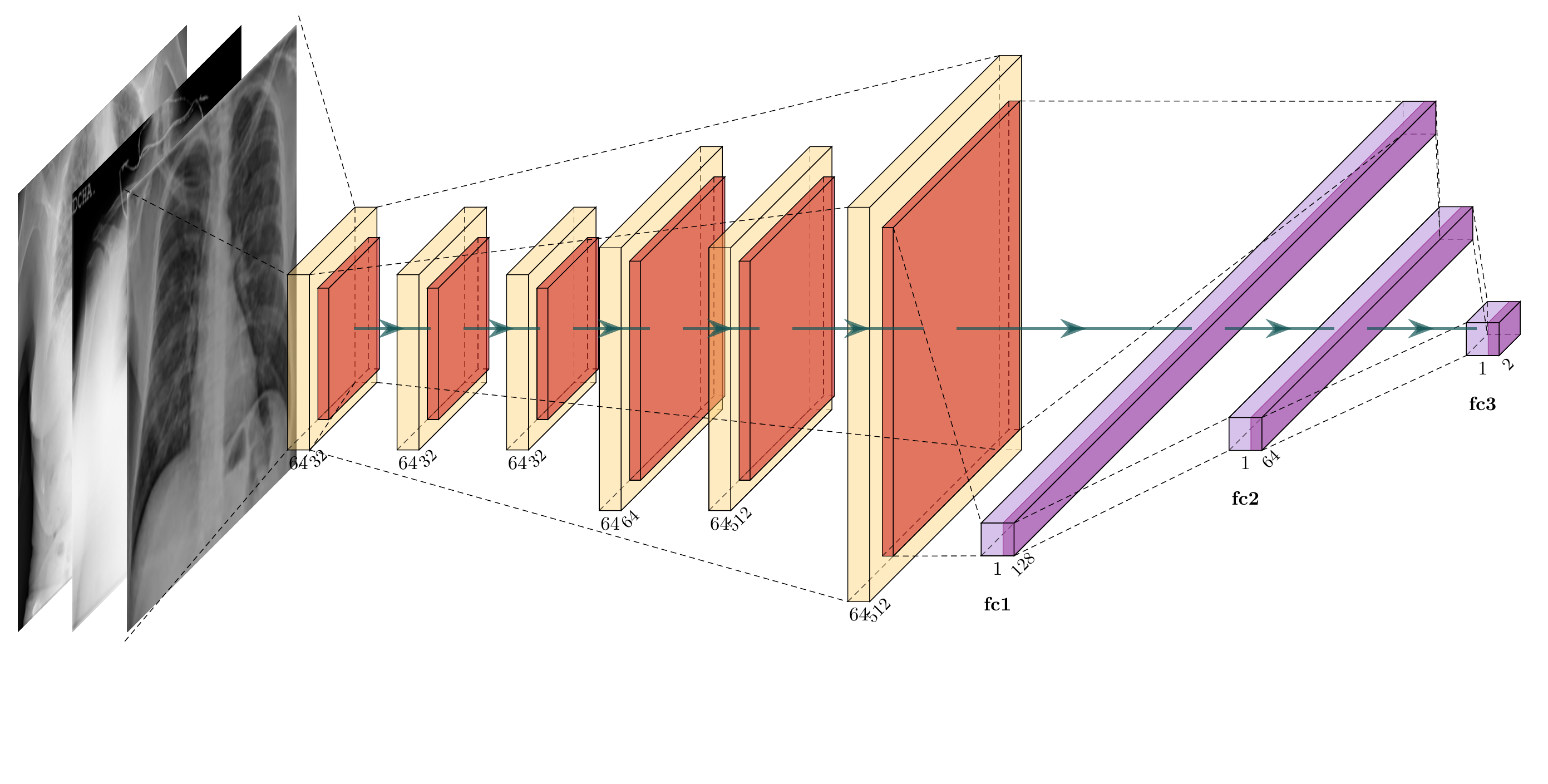}
    \caption{CNN based COVID-19 detection model.}
    \label{fig:cnn_model}
\end{figure}

Figure \ref{fig:cnn_model} shows CNN based COVID-19 detection model used in the experiments.

\subsection{Notations}

\begin{itemize}
    \item Boldface lowercase letters show the vectors (e.g., $\mathbf{x}$)
    \item $\llbracket W \rrbracket$ shows the ciphertext of a matrix $W$.
    \item $\oplus$ shows homomorphic encryption based addition, $\otimes$ homomorphic encryption based multiplication.
    \item $(key_{pub}, key_{priv})$ shows public/private key pairs. 
\end{itemize}

\subsection{Client Initialization}

Algorithm \ref{alg:init} shows the overall process in the initialization phase. Each client trains the local classifier, $h_i$ with their private datase, $\mathcal{D}_i$. The trained model's weight matrix, $W$, is encrypted, $\llbracket W \rrbracket$, and shared with the server  

\begin{algorithm}[!htbp]
\caption{Model training in each client}\label{alg:init} \small
\begin{algorithmic}[1]
\Require The dataset at client $c$: $\mathcal{D}_{c} =\{(\mathbf{x},y) | \mathbf{x} \in \mathbb{R}^{m}, y \in \mathbb{R}\}_{i=0}^m$, public key: $Key_{pub}$
\State $X_{train},X_{test}, \mathbf{y}_{train}, \mathbf{y}_{test}  \gets train\_test\_split(\mathcal{D})$ 

\State $h \gets global\_model$
\State $h.fit(X_{train},\mathbf{y}_{train})$

\State $W \gets \emptyset$ \LineComment{Create an empty matrix for the encrypted layer weights}
\ForEach{$layer \in h$}
   \State $\llbracket W \rrbracket \gets encrypt\_fractional(layer.weights, key_{pub})$ \LineComment{Encrypt the layer weights ($layer.weights \in \mathbb{R}^m$) with public key.} 
\EndFor
\State \textbf{Return} $\llbracket W \rrbracket$ \LineComment{The encrypted weight matrix}
\end{algorithmic}
\end{algorithm}

\subsection{Model Aggregation}

The server collects all encrypted weight matrices, $\{\llbracket W \rrbracket_0, \cdots, \llbracket W \rrbracket_c\}$, from the clients. It calculates the average weight value of each neuron in the encrypted domain. Algorithm \ref{alg:aggr} shows the overall process in the aggregation phase.

\begin{algorithm}[!htbp]
\caption{Model aggregation at the server}\label{alg:aggr} \small
\begin{algorithmic}[1]
\Require public key: $Key_{pub}$, the number of clients: $c$, client model weights: $H = \{\llbracket W \rrbracket_i\}_{i=0}^c$
\State $\llbracket W \rrbracket_{aggr} \gets \emptyset$
\ForEach{$h \in H$}
   \ForEach{$\llbracket row \rrbracket \in h$}
      \State $\llbracket W \rrbracket_{aggr} \gets \llbracket W \rrbracket_{aggr} \oplus \llbracket row \rrbracket$ \LineComment{Homomorphic addition}
   \EndFor
\EndFor
\ForEach{$\llbracket row \rrbracket \in \llbracket W \rrbracket_{aggr}$}
   \State $\llbracket row \rrbracket \gets \llbracket row \rrbracket \otimes c^{-1}$ \LineComment{Homomorphic multiplication.}
\EndFor
\State \textbf{Return} $\llbracket W \rrbracket_{aggr}$ \LineComment{Return the aggregated weight matrix in the encrypted domain}
\end{algorithmic}
\end{algorithm}

\subsection{Client Decryption}
The last step is client decryption which each client decrypt the aggregated and encrypted weight matrix, $\llbracket W \rrbracket_{aggr}$, and updates their local model, $h$. Algorithm \ref{alg:decryption} shows the overall process in the client decryption phase.

\begin{algorithm}[!htbp]
\caption{Client decryption}\label{alg:decryption} \small
\begin{algorithmic}[1]
\Require private key: $Key_{priv}$, encrypted aggregated weights: $\llbracket W \rrbracket_{aggr}$
\State $h \gets global\_model$
\ForEach{$layer \in h$}
   \State $\llbracket row \rrbracket \gets \llbracket W \rrbracket_{aggr}(layer)$ \LineComment{Get the corresponding row for layer}
   \State $layer \gets decrypt\_fractional(\llbracket row \rrbracket, key_{priv})$ \LineComment{Decrypt the row and update the layer weights}
\EndFor
\State $h.save\_model(global\_model)$ \LineComment{Save the aggregated model as global\_model at client. }
\end{algorithmic}
\end{algorithm}

\section{Results}
\subsection{Experimental Setup}
We have implemented our proposed protocols and the classifier training phase in Python by using the Keras/Tensorflow libraries for the model building and the Microsoft SEAL library for the somewhat homomorphic encryption implementation. To show the training phase time performance of the proposed protocols, we tested COVID-19 x-ray scans public dataset with different number of clients and the ciphertext modulus, $q=\{128,192\}$, which determines how much noise can accumulate before decryption fails. Table \ref{tab:ds_desc} shows the dataset details.

\begin{table}[!htbp]
    \centering
    \caption{Dataset description}
    \label{tab:ds_desc}
    \begin{tabular}{|c|c|c|}
    \hline
        \textbf{Dataset} & \textbf{Rows} & \textbf{Label} \\ \hline \hline
        \multirow{2}{*}{\textbf{Training}}& 800 & Negative \\
        & 800 & Positive \\ \hline
        \multirow{2}{*}{\textbf{Test}} & 200 & Negative \\
        & 200 & Positive \\ \hline
    \end{tabular}
\end{table}

\begin{figure}
     \centering
     \begin{subfigure}[b]{0.32\linewidth}
         \centering
         \includegraphics[width=\linewidth]{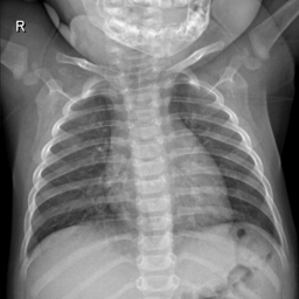}
         \caption{}
         \label{fig:covid-1}
     \end{subfigure}
     \hfill
     \begin{subfigure}[b]{0.32\linewidth}
         \centering
         \includegraphics[width=\linewidth]{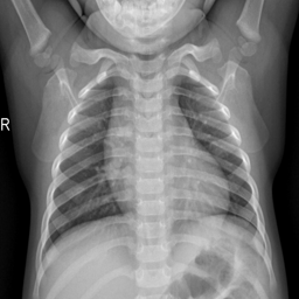}
         \caption{}
         \label{fig:covid-2}
     \end{subfigure}
     \hfill
     \begin{subfigure}[b]{0.32\linewidth}
         \centering
         \includegraphics[width=\linewidth]{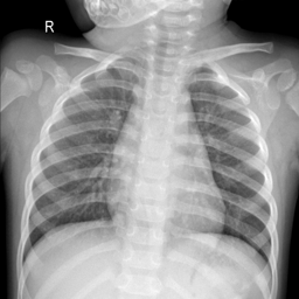}
         \caption{}
         \label{fig:covid-3}
     \end{subfigure}
     \begin{subfigure}[b]{0.32\linewidth}
         \centering
         \includegraphics[width=\linewidth]{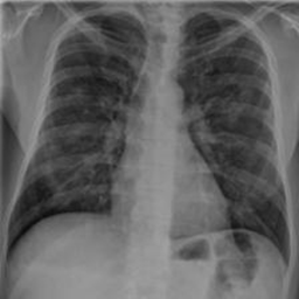}
         \caption{}
         \label{fig:covid-4}
     \end{subfigure}
     \hfill
     \begin{subfigure}[b]{0.32\linewidth}
         \centering
         \includegraphics[width=\linewidth]{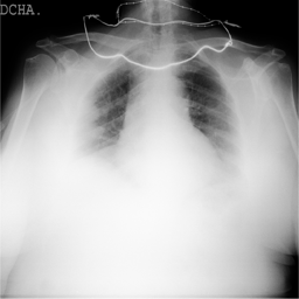}
         \caption{}
         \label{fig:covid-5}
     \end{subfigure}
     \hfill
     \begin{subfigure}[b]{0.32\linewidth}
         \centering
         \includegraphics[width=\linewidth]{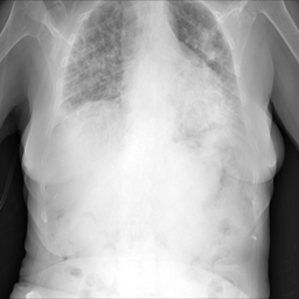}
         \caption{}
         \label{fig:covid-6}
     \end{subfigure}
        \caption{An example of an X-ray scan images taken from the dataset (a, b, c) with a label of COVID-19 negative, (d, e, f) COVID-19 positive.}
        \label{fig:dataset_sample}
\end{figure}

Samples of the dataset are depicted in Figure \ref{fig:dataset_sample}.

The dataset is arbitrarily partitioned
among each client ($c \in \{2,3,5,7\}$). , and then the prediction performance results in the encrypted-domain are compared with the results of the plain-domain.

\subsection{Experimental Results}

Table \ref{tab:init} shows the best performance of the conventional CNN method of COVID-19 Xray scans dataset.

\begin{table}[htbp!]
    \centering
    \caption{Initial results in plain domain without using federated learning}
    \label{tab:init}
    \begin{tabular}{|c|c|}
    \hline
    \textbf{Metric} & \textbf{Value} \\ \hline \hline
         Precision & 0.868924 \\
Recall    & 0.840000 \\
F1 Score  & 0.836801 \\
Accuracy  & 0.840000 \\ \hline
    \end{tabular}
\end{table}

Table \ref{tab:pred_results} shows the prediction performance of the CNN based classification model with and without encryption. As shown in the table, when the number of clients
varies from 2 to 7, then the overall prediction performance stays relatively stable at about 0.84 in the proposed training method.

\begin{table*}[!htbp]
    \centering
    \caption{Prediction performance of the somewhat HE and plain numbers based federated learning models.}
    \label{tab:pred_results}
    \begin{tabular}{|c||rrr|rrr|rrr|rrr|}
        \hline
        \multirow{2}{*}{\textbf{Clients}} & \multicolumn{3}{|c|}{\textbf{Accuracy}} &\multicolumn{3}{c|}{\textbf{F1}} & \multicolumn{3}{c|}{\textbf{Precision}} & \multicolumn{3}{c|}{\textbf{Recall}} \\ \cline{2-13}
         &\textbf{128} &  \textbf{192} &  \textbf{Plain} &  \textbf{128} & \textbf{192} &  \textbf{Plain} &  \textbf{128} &  \textbf{192} &  \textbf{Plain} &  \textbf{128} &  \textbf{192} &  \textbf{Plain} \\
        \hline \hline
         2 & 0.8375 &   0.8400 &     0.8450 & 0.834132 & 0.837030 &  0.842123 &  0.867337 &  0.866735 &    0.872128 &   0.8375 &   0.8400 &     0.8450 \\
         3 &0.8400 &   0.8400 &     0.8375 & 0.838040 & 0.836801 &  0.834369 &  0.857293 &  0.868924 &    0.865112 &   0.8400 &   0.8400 &     0.8375 \\
         5 &0.8300 &   0.8325 &     0.8350 & 0.827078 & 0.829732 &  0.832164 &  0.853925 &  0.855624 &    0.859288 &   0.8300 &   0.8325 &     0.8350 \\
         7 &0.8525 &   0.8450 &     0.8275 & 0.850776 & 0.842540 &  0.824649 &  0.869584 &  0.868000 &    0.850277 &   0.8525 &   0.8450 &     0.8275 \\
        \hline
    \end{tabular}
\end{table*}

\begin{figure}[!htbp]
    \centering
    \includegraphics[width=1.0\linewidth]{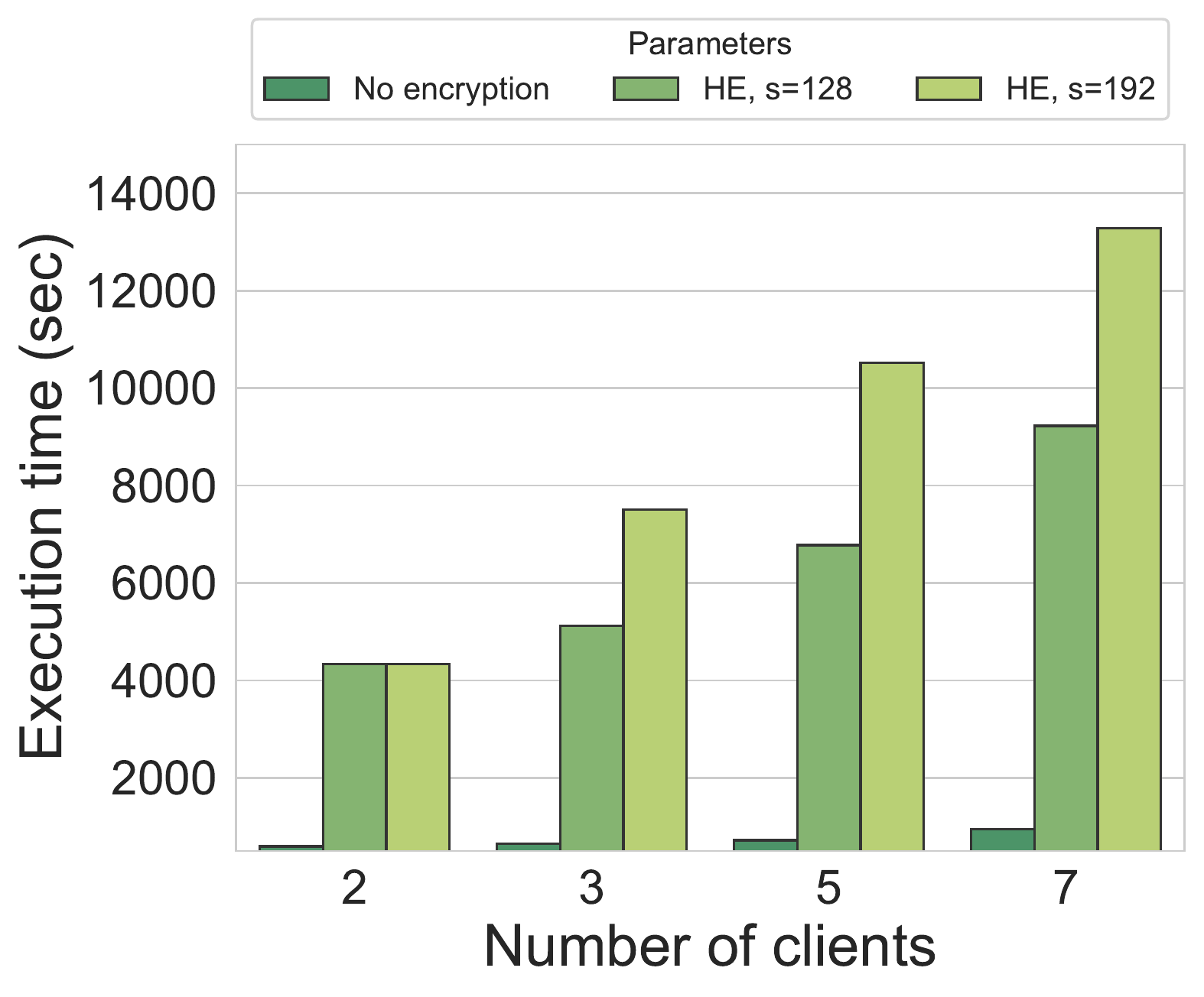}
    \caption{Execution time in seconds with the different security levels.}
    \label{fig:exec_time}
\end{figure}

Figure \ref{fig:exec_time} shows the execution times in seconds with three different configuration (i.e. plain, s=128, s=192). As expected, the execution in the encrypted domain is much higher than the plain domain.

\section{Discussion}

The experimental results in figure \ref{fig:exec_time} provides new insights into the relationships between different number of clients and execution time. There is a significant difference in execution time between plain ( Unencrypted) and encrypted data processes. This exponential differences are due to the complexity of the homomorphic encryption and processing encrypted data. However the execution times of different ciphertext modulus values (128,192) are indistinguishable for two clients but, execution time variation is rising with the growing the number of clients. That being so, there is an anticipated trade off between execution time and security level of the models. 

For the prediction phase, the test performances of the both encrypted and unencrypted processes are very similar as indicated in table \ref{tab:pred_results}. In fact, similar performances are achieved by each model with increasing the number of clients. Moreover, for some cases, results with plain data performs slightly better than the applied encryption results. For instance, the accuracy results of five clients indicates that plain versions accomplished better for each metric namely, accuracy, F1, precision, and Recall.

\section{Conclusion}

Privacy preserving become an essential practice of healthcare institutions as it is mandated by both EU and the US. Federated learning and homomorphic encryption will play critical role to maintain data security and model training. With benefitting from both techniques, the proposed model achieves compatitive performance while there is a significant trade off for the execution time and number of clients. The classification metrics, i.e. accuracy, F1. precision and recall, reaches over \%80 using both encrypted and plain data for each federated learning case. 

The privacy attacks will cause immense damages to the security and privacy of the patient information. This will hinder the advancement in healthcare using data-driven models. Therefore it is indispensable to take imperative steps to strengthen not only the safety of the information but also the way data is processed. This study demonstrated that federated learning with homomorphic encryption could be successfully applied to enhance data-driven models by eliminating and minimizing the share of the sensitive data. It is envisioned that this study could be useful for the scientists and researchers working on the sensitive healthcare data in multi-party computation settings.

\bibliographystyle{ACM-Reference-Format}
\bibliography{sample-base}


\begin{thebibliography}{25}


\ifx \showCODEN    \undefined \def \showCODEN     #1{\unskip}     \fi
\ifx \showDOI      \undefined \def \showDOI       #1{#1}\fi
\ifx \showISBNx    \undefined \def \showISBNx     #1{\unskip}     \fi
\ifx \showISBNxiii \undefined \def \showISBNxiii  #1{\unskip}     \fi
\ifx \showISSN     \undefined \def \showISSN      #1{\unskip}     \fi
\ifx \showLCCN     \undefined \def \showLCCN      #1{\unskip}     \fi
\ifx \shownote     \undefined \def \shownote      #1{#1}          \fi
\ifx \showarticletitle \undefined \def \showarticletitle #1{#1}   \fi
\ifx \showURL      \undefined \def \showURL       {\relax}        \fi
\providecommand\bibfield[2]{#2}
\providecommand\bibinfo[2]{#2}
\providecommand\natexlab[1]{#1}
\providecommand\showeprint[2][]{arXiv:#2}

\bibitem[Abouelmehdi et~al\mbox{.}(2017)]%
        {abouelmehdi2017big}
\bibfield{author}{\bibinfo{person}{Karim Abouelmehdi},
  \bibinfo{person}{Abderrahim Beni-Hssane}, \bibinfo{person}{Hayat Khaloufi},
  {and} \bibinfo{person}{Mostafa Saadi}.} \bibinfo{year}{2017}\natexlab{}.
\newblock \showarticletitle{Big data security and privacy in healthcare: A
  Review}.
\newblock \bibinfo{journal}{\emph{Procedia Computer Science}}
  \bibinfo{volume}{113} (\bibinfo{year}{2017}), \bibinfo{pages}{73--80}.
\newblock


\bibitem[Acar et~al\mbox{.}(2018)]%
        {acar2018survey}
\bibfield{author}{\bibinfo{person}{Abbas Acar}, \bibinfo{person}{Hidayet Aksu},
  \bibinfo{person}{A~Selcuk Uluagac}, {and} \bibinfo{person}{Mauro Conti}.}
  \bibinfo{year}{2018}\natexlab{}.
\newblock \showarticletitle{A survey on homomorphic encryption schemes: Theory
  and implementation}.
\newblock \bibinfo{journal}{\emph{ACM Computing Surveys (Csur)}}
  \bibinfo{volume}{51}, \bibinfo{number}{4} (\bibinfo{year}{2018}),
  \bibinfo{pages}{1--35}.
\newblock


\bibitem[Ali et~al\mbox{.}(2022)]%
        {ali2022deep}
\bibfield{author}{\bibinfo{person}{Aitizaz Ali},
  \bibinfo{person}{Muhammad~Fermi Pasha}, \bibinfo{person}{Jehad Ali},
  \bibinfo{person}{Ong~Huey Fang}, \bibinfo{person}{Mehedi Masud},
  \bibinfo{person}{Anca~Delia Jurcut}, {and} \bibinfo{person}{Mohammed~A
  Alzain}.} \bibinfo{year}{2022}\natexlab{}.
\newblock \showarticletitle{Deep Learning Based Homomorphic Secure Search-Able
  Encryption for Keyword Search in Blockchain Healthcare System: A Novel
  Approach to Cryptography}.
\newblock \bibinfo{journal}{\emph{Sensors}} \bibinfo{volume}{22},
  \bibinfo{number}{2} (\bibinfo{year}{2022}), \bibinfo{pages}{528}.
\newblock


\bibitem[Alloghani et~al\mbox{.}(2019)]%
        {alloghani2019systematic}
\bibfield{author}{\bibinfo{person}{Mohamed Alloghani},
  \bibinfo{person}{Mohammed~M Alani}, \bibinfo{person}{Dhiya Al-Jumeily},
  \bibinfo{person}{Thar Baker}, \bibinfo{person}{Jamila Mustafina},
  \bibinfo{person}{Abir Hussain}, {and} \bibinfo{person}{Ahmed~J Aljaaf}.}
  \bibinfo{year}{2019}\natexlab{}.
\newblock \showarticletitle{A systematic review on the status and progress of
  homomorphic encryption technologies}.
\newblock \bibinfo{journal}{\emph{Journal of Information Security and
  Applications}}  \bibinfo{volume}{48} (\bibinfo{year}{2019}),
  \bibinfo{pages}{102362}.
\newblock


\bibitem[Antunes et~al\mbox{.}(2022)]%
        {antunes2022federated}
\bibfield{author}{\bibinfo{person}{Rodolfo~Stoffel Antunes},
  \bibinfo{person}{Cristiano~Andr{\'e} da Costa}, \bibinfo{person}{Arne
  K{\"u}derle}, \bibinfo{person}{Imrana~Abdullahi Yari}, {and}
  \bibinfo{person}{Bj{\"o}rn Eskofier}.} \bibinfo{year}{2022}\natexlab{}.
\newblock \showarticletitle{Federated Learning for Healthcare: Systematic
  Review and Architecture Proposal}.
\newblock \bibinfo{journal}{\emph{ACM Transactions on Intelligent Systems and
  Technology (TIST)}} (\bibinfo{year}{2022}).
\newblock


\bibitem[Barbieri et~al\mbox{.}(2022)]%
        {barbieri2022decentralized}
\bibfield{author}{\bibinfo{person}{Luca Barbieri}, \bibinfo{person}{Stefano
  Savazzi}, \bibinfo{person}{Mattia Brambilla}, {and} \bibinfo{person}{Monica
  Nicoli}.} \bibinfo{year}{2022}\natexlab{}.
\newblock \showarticletitle{Decentralized federated learning for extended
  sensing in 6G connected vehicles}.
\newblock \bibinfo{journal}{\emph{Vehicular Communications}}
  \bibinfo{volume}{33} (\bibinfo{year}{2022}), \bibinfo{pages}{100396}.
\newblock


\bibitem[Bocu and Costache(2018)]%
        {bocu2018homomorphic}
\bibfield{author}{\bibinfo{person}{Razvan Bocu} {and} \bibinfo{person}{Cosmin
  Costache}.} \bibinfo{year}{2018}\natexlab{}.
\newblock \showarticletitle{A homomorphic encryption-based system for securely
  managing personal health metrics data}.
\newblock \bibinfo{journal}{\emph{IBM Journal of Research and Development}}
  \bibinfo{volume}{62}, \bibinfo{number}{1} (\bibinfo{year}{2018}),
  \bibinfo{pages}{1--1}.
\newblock


\bibitem[Brakerski(2012)]%
        {10.1007/978-3-642-32009-5_50}
\bibfield{author}{\bibinfo{person}{Zvika Brakerski}.}
  \bibinfo{year}{2012}\natexlab{}.
\newblock \showarticletitle{Fully Homomorphic Encryption without Modulus
  Switching from Classical GapSVP}. In \bibinfo{booktitle}{\emph{Advances in
  Cryptology -- CRYPTO 2012}}, \bibfield{editor}{\bibinfo{person}{Reihaneh
  Safavi-Naini} {and} \bibinfo{person}{Ran Canetti}} (Eds.).
  \bibinfo{publisher}{Springer Berlin Heidelberg}, \bibinfo{address}{Berlin,
  Heidelberg}, \bibinfo{pages}{868--886}.
\newblock
\showISBNx{978-3-642-32009-5}


\bibitem[Catak et~al\mbox{.}(2020)]%
        {electronics9020229}
\bibfield{author}{\bibinfo{person}{Ferhat~Ozgur Catak}, \bibinfo{person}{Ismail
  Aydin}, \bibinfo{person}{Ogerta Elezaj}, {and} \bibinfo{person}{Sule
  Yildirim-Yayilgan}.} \bibinfo{year}{2020}\natexlab{}.
\newblock \showarticletitle{Practical Implementation of Privacy Preserving
  Clustering Methods Using a Partially Homomorphic Encryption Algorithm}.
\newblock \bibinfo{journal}{\emph{Electronics}} \bibinfo{volume}{9},
  \bibinfo{number}{2} (\bibinfo{year}{2020}).
\newblock
\showISSN{2079-9292}
\urldef\tempurl%
\url{https://doi.org/10.3390/electronics9020229}
\showDOI{\tempurl}


\bibitem[{{\c{S}}ahinba{\c{s}}} and {Ozgur Catak}(2021)]%
        {csahinbacs2021secure}
\bibfield{author}{\bibinfo{person}{Kevser {{\c{S}}ahinba{\c{s}}}} {and}
  \bibinfo{person}{Ferhat {Ozgur Catak}}.} \bibinfo{year}{2021}\natexlab{}.
\newblock \showarticletitle{{Secure Multi-Party Computation based Privacy
  Preserving Data Analysis in Healthcare IoT Systems}}.
\newblock \bibinfo{journal}{\emph{arXiv e-prints}}, Article
  \bibinfo{articleno}{arXiv:2109.14334} (\bibinfo{date}{Sept.}
  \bibinfo{year}{2021}), \bibinfo{numpages}{arXiv:2109.14334}~pages.
\newblock
\showeprint[arxiv]{2109.14334}~[cs.CR]


\bibitem[Fan and Vercauteren(2012)]%
        {Fan2012SomewhatPF}
\bibfield{author}{\bibinfo{person}{Junfeng Fan} {and} \bibinfo{person}{Frederik
  Vercauteren}.} \bibinfo{year}{2012}\natexlab{}.
\newblock \showarticletitle{Somewhat Practical Fully Homomorphic Encryption}.
\newblock \bibinfo{journal}{\emph{IACR Cryptol. ePrint Arch.}}
  \bibinfo{volume}{2012} (\bibinfo{year}{2012}), \bibinfo{pages}{144}.
\newblock


\bibitem[Ibarrondo and Viand(2021)]%
        {EURECOM_6801}
\bibfield{author}{\bibinfo{person}{Alberto Ibarrondo} {and}
  \bibinfo{person}{Alexander Viand}.} \bibinfo{year}{2021}\natexlab{}.
\newblock \showarticletitle{Pyfhel: Python for homomorphic encryption
  libraries}. In \bibinfo{booktitle}{\emph{WAHC 2021, 9th Workshop on Encrypted
  Computing \&amp; Applied Homomorphic Cryptography, Associated with the ACM
  CCS 2021 conference, 15 November 2021, Seoul, South Korea}},
  \bibfield{editor}{\bibinfo{person}{ACM}} (Ed.). \bibinfo{address}{Seoul}.
\newblock


\bibitem[Kairouz et~al\mbox{.}(2021)]%
        {kairouz2021advances}
\bibfield{author}{\bibinfo{person}{Peter Kairouz}, \bibinfo{person}{H~Brendan
  McMahan}, \bibinfo{person}{Brendan Avent}, \bibinfo{person}{Aur{\'e}lien
  Bellet}, \bibinfo{person}{Mehdi Bennis}, \bibinfo{person}{Arjun~Nitin
  Bhagoji}, \bibinfo{person}{Kallista Bonawitz}, \bibinfo{person}{Zachary
  Charles}, \bibinfo{person}{Graham Cormode}, \bibinfo{person}{Rachel
  Cummings}, {et~al\mbox{.}}} \bibinfo{year}{2021}\natexlab{}.
\newblock \showarticletitle{Advances and open problems in federated learning}.
\newblock \bibinfo{journal}{\emph{Foundations and Trends{\textregistered} in
  Machine Learning}} \bibinfo{volume}{14}, \bibinfo{number}{1--2}
  (\bibinfo{year}{2021}), \bibinfo{pages}{1--210}.
\newblock


\bibitem[Kara et~al\mbox{.}(2021)]%
        {kara2021fully}
\bibfield{author}{\bibinfo{person}{Mostefa Kara}, \bibinfo{person}{Abdelkader
  Laouid}, \bibinfo{person}{Mohammed~Amine Yagoub}, \bibinfo{person}{Reinhardt
  Euler}, \bibinfo{person}{Saci Medileh}, \bibinfo{person}{Mohammad Hammoudeh},
  \bibinfo{person}{Amna Eleyan}, {and} \bibinfo{person}{Ahc{\`e}ne Bounceur}.}
  \bibinfo{year}{2021}\natexlab{}.
\newblock \showarticletitle{A fully homomorphic encryption based on magic
  number fragmentation and El-Gamal encryption: Smart healthcare use case}.
\newblock \bibinfo{journal}{\emph{Expert Systems}} (\bibinfo{year}{2021}),
  \bibinfo{pages}{e12767}.
\newblock


\bibitem[Kumar et~al\mbox{.}(2020)]%
        {kumar2020secure}
\bibfield{author}{\bibinfo{person}{A~Vijaya Kumar},
  \bibinfo{person}{Mogalapalli~Sai Sujith}, \bibinfo{person}{Kosuri~Tarun Sai},
  \bibinfo{person}{Galla Rajesh}, {and} \bibinfo{person}{Devulapalli
  Jagannadha~Sriram Yashwanth}.} \bibinfo{year}{2020}\natexlab{}.
\newblock \showarticletitle{Secure Multiparty computation enabled E-Healthcare
  system with Homomorphic encryption}. In \bibinfo{booktitle}{\emph{IOP
  Conference Series: Materials Science and Engineering}},
  Vol.~\bibinfo{volume}{981}. IOP Publishing, \bibinfo{pages}{022079}.
\newblock


\bibitem[Li et~al\mbox{.}(2019)]%
        {li2019privacy}
\bibfield{author}{\bibinfo{person}{Wenqi Li}, \bibinfo{person}{Fausto
  Milletar{\`\i}}, \bibinfo{person}{Daguang Xu}, \bibinfo{person}{Nicola
  Rieke}, \bibinfo{person}{Jonny Hancox}, \bibinfo{person}{Wentao Zhu},
  \bibinfo{person}{Maximilian Baust}, \bibinfo{person}{Yan Cheng},
  \bibinfo{person}{S{\'e}bastien Ourselin}, \bibinfo{person}{M~Jorge Cardoso},
  {et~al\mbox{.}}} \bibinfo{year}{2019}\natexlab{}.
\newblock \showarticletitle{Privacy-preserving federated brain tumour
  segmentation}. In \bibinfo{booktitle}{\emph{International workshop on machine
  learning in medical imaging}}. Springer, \bibinfo{pages}{133--141}.
\newblock


\bibitem[Rieke et~al\mbox{.}(2020)]%
        {rieke2020future}
\bibfield{author}{\bibinfo{person}{Nicola Rieke}, \bibinfo{person}{Jonny
  Hancox}, \bibinfo{person}{Wenqi Li}, \bibinfo{person}{Fausto Milletari},
  \bibinfo{person}{Holger~R Roth}, \bibinfo{person}{Shadi Albarqouni},
  \bibinfo{person}{Spyridon Bakas}, \bibinfo{person}{Mathieu~N Galtier},
  \bibinfo{person}{Bennett~A Landman}, \bibinfo{person}{Klaus Maier-Hein},
  {et~al\mbox{.}}} \bibinfo{year}{2020}\natexlab{}.
\newblock \showarticletitle{The future of digital health with federated
  learning}.
\newblock \bibinfo{journal}{\emph{NPJ digital medicine}} \bibinfo{volume}{3},
  \bibinfo{number}{1} (\bibinfo{year}{2020}), \bibinfo{pages}{1--7}.
\newblock


\bibitem[Roy et~al\mbox{.}(2019)]%
        {roy2019braintorrent}
\bibfield{author}{\bibinfo{person}{Abhijit~Guha Roy}, \bibinfo{person}{Shayan
  Siddiqui}, \bibinfo{person}{Sebastian P{\"o}lsterl}, \bibinfo{person}{Nassir
  Navab}, {and} \bibinfo{person}{Christian Wachinger}.}
  \bibinfo{year}{2019}\natexlab{}.
\newblock \showarticletitle{Braintorrent: A peer-to-peer environment for
  decentralized federated learning}.
\newblock \bibinfo{journal}{\emph{arXiv preprint arXiv:1905.06731}}
  (\bibinfo{year}{2019}).
\newblock


\bibitem[Sheller et~al\mbox{.}(2018)]%
        {sheller2018multi}
\bibfield{author}{\bibinfo{person}{Micah~J Sheller}, \bibinfo{person}{G~Anthony
  Reina}, \bibinfo{person}{Brandon Edwards}, \bibinfo{person}{Jason Martin},
  {and} \bibinfo{person}{Spyridon Bakas}.} \bibinfo{year}{2018}\natexlab{}.
\newblock \showarticletitle{Multi-institutional deep learning modeling without
  sharing patient data: A feasibility study on brain tumor segmentation}. In
  \bibinfo{booktitle}{\emph{International MICCAI Brainlesion Workshop}}.
  Springer, \bibinfo{pages}{92--104}.
\newblock


\bibitem[Talpur et~al\mbox{.}(2015)]%
        {talpur2015shared}
\bibfield{author}{\bibinfo{person}{Mir Sajjad~Hussain Talpur},
  \bibinfo{person}{Md~Zakirul~Alam Bhuiyan}, {and} \bibinfo{person}{Guojun
  Wang}.} \bibinfo{year}{2015}\natexlab{}.
\newblock \showarticletitle{Shared--node IoT network architecture with
  ubiquitous homomorphic encryption for healthcare monitoring}.
\newblock \bibinfo{journal}{\emph{International Journal of Embedded Systems}}
  \bibinfo{volume}{7}, \bibinfo{number}{1} (\bibinfo{year}{2015}),
  \bibinfo{pages}{43--54}.
\newblock


\bibitem[Tan et~al\mbox{.}(2020)]%
        {tan2020practical}
\bibfield{author}{\bibinfo{person}{Haowen Tan}, \bibinfo{person}{Pankoo Kim},
  {and} \bibinfo{person}{Ilyong Chung}.} \bibinfo{year}{2020}\natexlab{}.
\newblock \showarticletitle{Practical homomorphic authentication in
  cloud-assisted vanets with blockchain-based healthcare monitoring for
  pandemic control}.
\newblock \bibinfo{journal}{\emph{Electronics}} \bibinfo{volume}{9},
  \bibinfo{number}{10} (\bibinfo{year}{2020}), \bibinfo{pages}{1683}.
\newblock


\bibitem[{van Veen}(2018)]%
        {VANVEEN201870}
\bibfield{author}{\bibinfo{person}{Evert-Ben {van Veen}}.}
  \bibinfo{year}{2018}\natexlab{}.
\newblock \showarticletitle{Observational health research in Europe:
  understanding the General Data Protection Regulation and underlying debate}.
\newblock \bibinfo{journal}{\emph{European Journal of Cancer}}
  \bibinfo{volume}{104} (\bibinfo{year}{2018}), \bibinfo{pages}{70--80}.
\newblock
\showISSN{0959-8049}
\urldef\tempurl%
\url{https://doi.org/10.1016/j.ejca.2018.09.032}
\showDOI{\tempurl}


\bibitem[Wang and Zhang(2015)]%
        {wang2015data}
\bibfield{author}{\bibinfo{person}{Xiaoni Wang} {and}
  \bibinfo{person}{Zhenjiang Zhang}.} \bibinfo{year}{2015}\natexlab{}.
\newblock \showarticletitle{Data division scheme based on homomorphic
  encryption in WSNs for health care}.
\newblock \bibinfo{journal}{\emph{Journal of medical systems}}
  \bibinfo{volume}{39}, \bibinfo{number}{12} (\bibinfo{year}{2015}),
  \bibinfo{pages}{1--7}.
\newblock


\bibitem[Xu et~al\mbox{.}(2021)]%
        {xu2021federated}
\bibfield{author}{\bibinfo{person}{Jie Xu}, \bibinfo{person}{Benjamin~S
  Glicksberg}, \bibinfo{person}{Chang Su}, \bibinfo{person}{Peter Walker},
  \bibinfo{person}{Jiang Bian}, {and} \bibinfo{person}{Fei Wang}.}
  \bibinfo{year}{2021}\natexlab{}.
\newblock \showarticletitle{Federated learning for healthcare informatics}.
\newblock \bibinfo{journal}{\emph{Journal of Healthcare Informatics Research}}
  \bibinfo{volume}{5}, \bibinfo{number}{1} (\bibinfo{year}{2021}),
  \bibinfo{pages}{1--19}.
\newblock


\bibitem[Özgür Çatak and Mustacoglu(2018)]%
        {catak2018}
\bibfield{author}{\bibinfo{person}{Ferhat Özgür Çatak} {and}
  \bibinfo{person}{Ahmet~Fatih Mustacoglu}.} \bibinfo{year}{2018}\natexlab{}.
\newblock \showarticletitle{CPP-ELM: Cryptographically Privacy-Preserving
  Extreme Learning Machine for Cloud Systems}.
\newblock \bibinfo{journal}{\emph{International Journal of Computational
  Intelligence Systems}}  \bibinfo{volume}{11} (\bibinfo{year}{2018}),
  \bibinfo{pages}{33--44}.
\newblock
Issue 1.
\showISSN{1875-6883}
\urldef\tempurl%
\url{https://doi.org/10.2991/ijcis.11.1.3}
\showDOI{\tempurl}


\end{thebibliography}

\end{document}